\documentstyle[eqsecnum,aps,epsf,amssymb,prbbib]{revtex}

\def\tv{\tilde{v}}
\def\a{\alpha}
\def\b{\beta}

\def\g{\gamma}
\def\c{\chi}
\def\d{\delta}
\def\s{\sigma}
\def\D{\Delta}

\begin{document}
\draft

%\twocolumn[\hsize\textwidth\columnwidth\hsize\csname 
%@twocolumnfalse\endcsname

\title{Kinetics and thermodynamics across single-file pores:
solute permeability and rectified osmosis}

\author{Tom Chou\footnote{Present Address: MIT 
Mathematics Department, 77 Massachusetts Ave.,
Cambridge, MA 02139}}

\address{DAMTP and Dept. of Physiology,  
University of Cambridge, Cambridge CB3 9EW, ENGLAND}

%\address{$^{2}$Department of Physiology,  
%University of Cambridge, 
%Cambridge CB2 3EW, ENGLAND}

\date{\today}
\maketitle

\begin{abstract}
We study the effects of solute interactions on osmotic transport
through pores. By extending single-file, single-species kinetic
models to include entrance of solute into membrane pores, we model
the statistical mechanics of competitive transport of two species
across membrane pores. The results have direct applications to water
transport across biomembrane pores and particle movement in
zeolites, and can be extended to study ion channel transport.
Reflection coefficients, the reduction of osmotic fluxes measured
using different solutes, are computed in terms of the microscopic
kinetic parameters.  We find that a reduction in solvent flow due to
solute-pore interactions can be modelled by a Langmuir adsorption
isotherm. Osmosis experiments are discussed and proposed. Special
cases and Onsager relations are presented in the Appendices.
\end{abstract}

%\widetext

\section{Introduction}

Recent X-ray\cite{SCIENCE} and electron crystallographic
studies\cite{MITRA97,JAP} have yielded structures of
integral membrane proteins such as K$^{+}$ ion and water
channels to near atomic resolution.  Since many biological
transport channels have specificity in allowing specific
molecules to permeate and mediate simultaneous flows of
numerous species.\cite{FINK,ZEUTHEN} These new data may
offer insights that may help correlate structure to
function.  The pores spanning biomembranes as well as those found in
zeolites (important in numerous industrial processes such
as hydrocarbon separations and catalytic 
agents) are typically few Angstroms in diameter, and
contain very few molecules at any time. 
Therefore, we will study osmosis-driven 
particle transport across such pores using 
simple one-dimensional lattice models.

The next section reviews the linear phenomenological
expressions\cite{KAT1} and their parameters, allowing for a second
interacting, competing species. The empirical parameters arise from
macroscopic considerations only and are not derived from any
microscopic pore structure.  Section III formally defines a
two-species model similar to that used by Su {\it et al.}\cite{SU} and
Wang {\it et al.}\cite{2SITES} Here, we assume a microscopic structure
consisting of single-file pores. Unlike the single species case, where
the statistics of one-dimensional particle dynamics are well
understood,\cite{MILG1,ASEM,EVANS,PRL,TOBE,HAHN97} if solutes are
allowed to enter a pore, no general analytical solution exists for the
current across channels of arbitrary length. However, similar
assumptions implicit in modelling single-species, single-file
transport are used here: All nonlinearities except for those
associated with the internal particle dynamics are neglected.
Although our analysis is semiquantitative, and neglects some precise
details about the specific particle-particle and particle-pore
interactions, it provides a physically and mathematically consistent
microscopic picture of nonequilibrium flow through single-file, solute
permeable pores.

In the Solutions and Discussion section, we plot the behavior of
steady state flows under a variety of experimentally motivated
conditions.  Certain macroscopic conjectures of the role of solute
size on osmotic permeability\cite{HILLREV} are re-examined.  For short
pores that allow partial and total passage of solute, the effective
reflection coefficients (defined in Section II) and counterflow of
solute are also computed and plotted in a number of illustrative
graphs. An extension to longer pores is derived using an equilibrium
multi-site Langmuir adsorption model.\cite{HILL}

In the Conclusions, we summarize our analyses and discuss
interpretations of osmosis measurements.  We argue that models which
intuitively incorporate solute-membrane interactions into linear
coefficients\cite{HILLREV} are inconsistent with a virial expansion in
the solute-pore interactions.  Possible experiments and correlations among
the rate parameters are proposed.  For completeness, the lattice
statistics, rate equations for two-sectioned pores, Onsagers
relations, and analytic solutions in a special case are treated in
detail in the Appendices.

\section{Phenomenological Equations}

% **************************************
%
%   Figure 1, membrane and coefficients
%
\begin{figure}
\begin{center}
\leavevmode
\epsfxsize=3.2in
\epsfbox{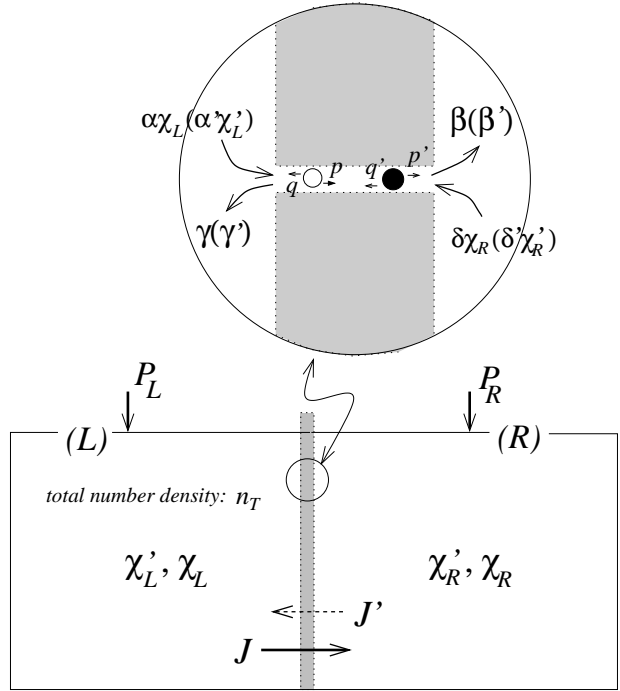}
\end{center}
\caption{Schematic of osmosis and pressure driven
flow through a membrane pore system. The reservoirs
$(L)$ and $(R)$ are assumed infinite.  The
coefficients in the inset $\{\xi\}=(\a,\b,\g,\d, p,
q)$, and $\{\xi'\} =(\a', \b', \g', \d', p', q')$
are microscopic conditional transport rates
for solvent and solute respectively and are defined by the 
figure.  The
solvent(solute) mole fractions in the $(L)$ and
$(R)$ reservoirs are $\c_{L}(\c'_{L})$ and
$\c_{R}(\c'_{R})$ respectively, and
$\c_{L,R}+\c'_{L,R} = 1$. In a typical osmosis
experiment, the membrane is impermeable to solute,
({\it e.g.} $\gamma'$ or $\delta' = 0$), and the
hydrostatic pressure difference $\Delta P =
P_{R}-P_{L}=0$.}
\label{FIG1}
\end{figure}
%
%
% ************************************** 

%\subsection{Linear Models}

We first review the linear phenomenological equations describing two
species transport across a membrane separating two infinite reservoirs
(Fig. \ref{FIG1}).  When a membrane is permeable to solute as well as
solvent, their fluxes, $J'$ and $J$, are coupled.  Following
Katchalsky and Curran \cite{KAT1} and defining the fluxes in terms of
a volume flow $J_{V}\equiv \tilde{v}J+\tilde{v}'J'$ conjugate to
hydraulic pressure driving forces, and a relative flow $J_{D}\equiv
J'/n'_{L}-J/n_{L}$ conjugate to diffusive driving forces, a set of
linear Onsager relations are derived:

\begin{equation}
\begin{array}{ll}
J_{V} & = L_{P}\D P + L_{PD} \D\Pi \\[13pt]
J_{D} & = L_{DP}\D P + L_{D} \D\Pi.
\end{array}
\label{ONSAGER}
\end{equation}

\noindent Here, $\tv(\tv')$ is the molecular volume
of solvent(solute) in solution, $n_{L}(n'_{L})$ is the solvent(solute)
particle number density in reservoir $(L)$. The osmotic pressure $\D
\Pi \cong n_{T}k_{B}T \D\chi'$, where $n_{T}$ is the total particle
number density which we approximate as equal in the two reservoirs.
The condition for zero volume flow $(J_{V} = 0)$ requires $\D P =
-(L_{PD}/L_{P})\D\Pi$.  Thus, equilibrium across solute-impermeable
membranes requires $\Delta P = \Delta \Pi$, and the reflection
coefficient \cite{KAT1} $\Sigma \equiv -L_{PD}/L_{P} = 1$.  For
permeable solutes, the condition for zero volume flow yields a
reflection coefficient $\Sigma < 1$ implying that a smaller hydraulic
pressure is required to balance volume flow under a steady state
osmotic pressure $\D \Pi$,

\begin{equation}
J_{V} = L_{P}(\D P-\Sigma\D \Pi).
\end{equation}

\noindent Since $L_{PD} = -L_{P}\, (\Sigma = 1)$ for
an impermeable solute, $\Sigma$ can be measured by comparing $J_{V}$
induced by permeable solutes to the maximum ($J_{V} \equiv
\tilde{v}J_{max}$) induced by impermeable solutes. However, the
experimental measurement of

\begin{equation}
\Sigma = {J_{V} \over \tv J_{max}} = 
{\tilde{v}J + \tilde{v}'J' \over \tilde{v}J_{max}}
\label{SIGMA}
\end{equation} 

\noindent (under isobaric $\D P=0$ conditions) does
not rely on the linearity assumed by (\ref{ONSAGER}), and nonlinear
mechanisms may manifest themselves in $\Sigma$.  Furthermore, as we
will see, the
current derived from an impermeable solute can also depend (to higher
order in solute concentration difference $\D \chi'$) on how the solute
particles partly enter and bind to the pore interiors. Thus,
measurements of $\Sigma$ may have more intricate dependences on
solute-pore structure than is sometimes taken into account.

\section{Two Species 1D Exclusion Model}

We now study flow mediated by one-dimensional channels.  Single
species exclusion models\cite{MILG1,PRL,TOBE,HAHN97} are extended to
allow partial or complete entrance of solute particles inside a pore.
This approach qualitatively models noncylindrical shapes\cite{SOKOL92}
often found in biological pores and ion channels, {\it i.e.} flares,
conical sections \cite{SCIENCE,FINK}, and vestibules \cite{MITRA97} in
the pore structure.  These wider sections may allow solute entrance
and binding, and may render the pore permeable to solute.  Ion
specific channels are also believed to have a narrower section near
the midplane which acts like a selectivity filter for only those ions
which can bind or pass through.

%\subsection{Chain dynamics}
%\subsection{Rate Equations for Short Pores}
The one-dimensional chain shown in Fig. 1 which allows two types of
particles ({\it e.g.} solvent and solute, or A and B) to enter and
occupy any site.\cite{ASEM,HAHN97} The kinetic rates pictured in Fig.
1 are defined as follows. A solvent particle enters site $i=1$ from
$(L)$ with probability per unit time $dt$ or rate $\a\chi_{L}$ if and
only if site $i=1$ is empty.  This entrance rate is given by an
intrinsic rate $\a$ times the mole fraction of solute particles in
$(L)$.  Similarly, the entrance rate from $(R)$ into the empty site
$i=N$ is given by $\d\chi_{R}$.  The exit rates from the pore to the
reservoirs at site $i=1, N$ provided they are occupied are $\gamma,
\beta$ respectively.  In the pore interior, a solvent particle hops to
the left with probability $q$ only if the site $i-1$ to the left is
empty.  transitions to an empty site to the right are denoted by $p$.
If the particles do not experience pondermotive forces (electrostatic
or gravitational forces) and are being transported across microscopic,
uniform pores, $p\simeq q$. However,, in general, the pores need not
be uniform, and pondermotive forces (such as electric fields acting on
charged particles) may also exist such that $p\neq q$. Completely
analogous rates are defined for the second type of particle, or
solutes, by primed quantities. The chain length is $L=N\ell$, where
$\ell$ is set by the size of the larger of the two types of particles.
If $a$ and $a'$ are the approximate molecular sizes of the solvent and
solute molecules, $a\gtrsim a'/2$, then each site can contain at most
one particle of either type.
 
By considering transitions among all possible particle
configurations, the kinetic steps defined by
(\ref{JINTERNAL}) and (\ref{JEXTERNAL}) in Appendix A can
be written in terms of dynamical rate equations. 
When solutes cannot enter the pore ($\a'=\d'=0$),
the solvent current $J_{max}$ 
across symmetric pores ($p=q$) in the presence of
noninteracting solutes ({\it i.e.}
$\a' = \b'=\g'=\d'=p'=q'=0$) is exactly\cite{PRL,TOBE}

\begin{equation}
J_{max}(N) = {p(\a\chi_{L}\b-\g\d\chi_{R})
\over (N-1)(\a\chi_{L}+\g)(\b+\d\chi_{R})+p(\a\chi_{L}+\b+\g+\d\chi_{R})}.
\end{equation}

When the solute particles can enter the pore, no analytic expressions
exist for $J$, $J'$.  However, for illustration, consider a membrane
so thin that $N=1$, and only boundary entrance and exit rates are
relevant.  The single site within the membrane can be in only one of
three possible states: empty ($P_{1}$), solvent filled ($P_{2}$), and
solute filled ($P_{3}$). The probability fluxes are determined by the
rate equations

\begin{equation}
\begin{array}{ll}
\partial_{t}P_{1} & =
-\left(\a\c_{L}+\a'\c'_{L}+\d\c_{R}+\d'\c'_{R}\right)
P_{1}+(\beta+\gamma)P_{2}+
(\beta'+\gamma') P_{3} \\[13pt]
\partial_{t}P_{2} & = (\alpha\chi_{L}+\delta\chi_{R})
P_{1}-(\beta+\gamma)P_{2} \\[13pt]
\partial_{t}P_{3} & =
(\alpha'\chi_{L}'+\delta'\chi_{R}')
P_{1}-(\beta'+\g')P_{3}
\end{array}
\end{equation}

\noindent Upon imposing steady state
($\partial_{t}P_{i} = 0$), and normalization ($P_{1}+P_{2}+P_{3}\equiv
1$), the averaged occupations $\langle\s\rangle,\, \langle\s'\rangle$
can be found and used in $J(N=1) = \alpha\chi_{L}
(1-\langle\s\rangle-\langle\s'\rangle)-\gamma\langle\s\rangle$ to find
the steady state solvent current

\begin{equation}
\displaystyle J(N=1) =
{(\alpha\chi_{L}\beta-\gamma\delta\chi_{R})
(\beta'+\g')\over 
(\alpha\chi_{L}+\beta+\gamma+\delta\chi_{R})
(\beta'+\gamma')+(\alpha'\chi'_{L}+\delta'\chi'_{R})
(\beta+\gamma)},
\label{J1}
\end{equation}

\noindent The solute current $J'$ is given by interchanging $
\{\xi\}\equiv (\alpha, \beta, \gamma, \delta) \leftrightarrow
\{\xi'\}\equiv (\alpha',\beta',\gamma',\delta')$ and $\chi_{L,R}
\leftrightarrow \chi'_{L,R}$ in (\ref{J1}).  Note that $J(N=1)$ has
the same form as $J_{max}(N=1)$ appropriate for single-species
transport except for an additional term in the denominator,

\begin{equation}
{(\a'\chi'_{L}+\d'\chi'_{R})(\b+\g)\over
\b'+\g'},
\end{equation}

\noindent representing interference from states with
solute occupation, $\s'=1, (P_{3})$.
The reflection coefficient defined
for a single-site pore is thus

\begin{equation}
\begin{array}{l}
\displaystyle \Sigma(N=1) =  
\displaystyle {1 \over 
1+{(\a'\c'_{L}+\d'\c'_{R})(\b+\g) \over
(\a\c_{L}+\b+\g+\d\c_{R})(\b'+\g')}} 
+{\tv\over \tv'}{J'(N=1) \over J_{max}(N=1)}
%\\[15pt]
%\quad\quad + {(\a'\c'_{L}\b'-\g'\d'\c'_{R})\over
%(\a\c_{L}\b-\g\d\c_{R})} {\tv'/\tv \over
%{\b'+\g' \over \b+\g} + {\a'\c'_{L}+\d'\c'_{R} \over 
%\a\c_{L}+\b+\g+\d\c_{R}}},
\end{array}
\end{equation}

\noindent and represents {\it two} factors; reduction of solvent
flow (first term), and solute backflow (second term). 

The microscopic activation energies $E_{\xi}(E'_{\xi}),$ and the rate
parameters $\{\xi\} = (\a,\b,\g,\d,p,q)$ and 
$\{\xi'\}= (\a',\b',\g',\d',p',q')$, are governed by the
microscopic membrane-solvent(solute) interactions, 
and are approximated by the Arrenhius forms 
$\xi \simeq \xi_{0}\exp(-E_{\xi}/k_{B}T)$ and 
$\xi' \simeq \xi_{0}'\exp(-E'_{\xi}/k_{B}T)$.

\begin{figure}
\begin{center}
\leavevmode
\epsfxsize=2in
\epsfbox{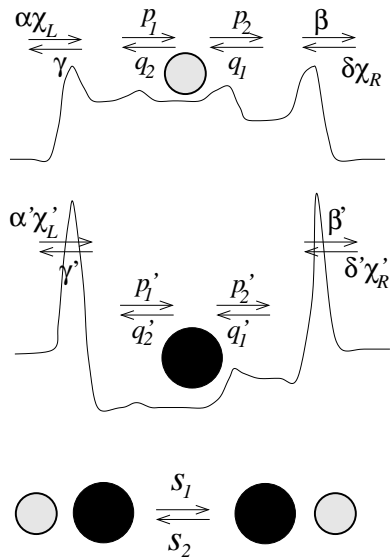}
\end{center}
\caption{Schematic of a three sectioned pore
displaying all microscopic energies and transition rates.
Separate energy landscapes and occur for solvent(open
circles) and solutes(filled circles). Exchange rates 
associated with wider pores are designated $s_{1}, _{2}$.
We take $s_{1}=s_{2}=s$ for the same reasons $q=p$.}
\label{FIG2}
\end{figure}

Kinetic equations can be readily solved for slightly
longer pores. The states and transition probabilities
for a two species, two section ($N=2$) pore are
enumerated in Appendix B. Currents across longer pores
become increasingly tedious to compute as the matrix
size increases as $3^{N}\times 3^{N}$.  The general
transition rates for an $N=3$ ($27 \times 27$ matrix)
pore are shown in Fig. \ref{FIG2} and will serve as
the basis for most our subsequent analyses.

% ********** Figure 2, 3-site energy landscape **********
%
%
% *******************************************************

\section{Solutions and Discussion}

Various effects of physical and chemical parameters, such
as pore length and solute interaction dependences, are
considered. We do not attempt to assign precise values to
$\{\xi\}, \{\xi'\}$, but take the simplest, physically
reasonable values to illustrate the relevant physical
mechanisms.  

\subsection{Solute Interacting, Impermeable Pores}

First consider solutes only in the $(R)$-reservoir that can
interact ({\it i.e.} partially enter) with a single-site
($N=1$) pore, but that encounters an infinite barrier while
on the left side of the site ({\it e.g.} $\d'\neq 0,
\g'=\c'_{L}=0$).  Assume for simplicity a pore that
remains microscopically symmetric with respect interactions
with the solvent.  A reduction of solvent flux manifests
itself in the definition of $\Sigma$ even though the pore
is impermeable to solute and $J'=0$.  Thus
$\Sigma(N=1,J'=0)$ reduces to

\begin{equation}
\begin{array}{l}
\displaystyle \Sigma_{\em eff}(N=1) \equiv {J(N=1)
\over J_{max}(N=1)} \quad  \\[13pt]
\displaystyle \: \quad \quad  =
\left[1+{2\d'\chi'_{R} \over 
(2(\bar{\a}+1)-\bar{\a}\chi'_{R})\b'}\right]^{-1} <
1
\label{SIGMAEFF}
\end{array}
\end{equation}

\noindent where $\bar{\a} \equiv \a/\b$ defines an
effective solvent-pore binding affinity.\cite{FOOT1}  We
find the reduction in osmotic solvent flux for solutes that
enter the site, represented by $\Sigma_{\em eff}$,
depends mainly on the solute equilibrium constant
$\d'\c'_{R}/\b'$, and $(\bar{\a}+1)^{-1}$. Since occupation
of the site by solute is required to influence (block)
solvent flow, the $\d'\c'_{R}/\b'$ behavior is not
surprising.  A low solvent affinity $\bar{\a}$ further
allows the site to be more likely occupied by solute. These
competitive factors determine the likelihood solute enters
the pore and hinders solvent flux.  Note that $\Sigma_{\em
eff}(\c'_{R}\rightarrow 0) \rightarrow 1$.

The dependence on solute concentration $\chi'_{R}$ is not unexpected.
Qualitatively, one power of $\chi'_{R}$ contributes the first order
driving force for osmosis (the entropy of mixing); solute-solute
(which we do not consider) and solute-membrane interactions must come
at higher orders of $\D \chi'$ via a virial expansion.  Theories that
suggest partial solute entrance into the pores \cite{HILLREV,MCGANN}
as a mechanism for reducing {\it solvent} flows require the presence
of solute in the vicinity (adsorbed near) the pore mouth. The
likelihood of this configuration will depend on the {\it bulk} solute
concentration. Thus, these types of nonlinearities can be effectively
built into a concentration dependent $P_{os}= P_{os}(\D
\chi')$.\cite{SOLOMON70} In the $\D\chi' \rightarrow 0$ limit,
$P_{os}$ approaches a constant (that depends only on how solvent
travels through the membrane pore). The lowest order term is 
independent of solute identity.

%\noindent The nonlinear terms in the expansion of
%$J(\d'=0,\g'=0,N=1) =d_{k}(\D\c')^{k}$ are important for
%solute mole fractions 

%\begin{equation}
%\D \chi' \gtrsim \arrowvert {d_{1} \over d_{2}}\arrowvert 
%= \arrowvert {2\b(\a+\b)+(2\a'\b-2\a\b'+\d')
%\c'_{L}\over \a\b'-2\b\d'}\arrowvert.
%\end{equation}

%\noindent Thus, nonlinear effects are more likely to obtain
%when the pore is solute attracting and $\d'/\b'$ is large
%($E_{\b}'\ll k_{B}T$), consistent with the conditions for
%$\Sigma_{\em eff} < 1$.

Equation (\ref{SIGMAEFF}) arises from a Langmuir adsorption
isotherm\cite{HILL} determining the statistical fraction of
solute-free (thus solvent conducting) states:

\begin{equation}
\Sigma_{\em eff} \simeq 1-P_{eq}\{\s' = 1\} 
\simeq {1+\a\c_{L}/\b \over Z}
\label{SIGMAAPPROX1}
\end{equation}

\noindent where $Z \propto  1 + \a\c_{L}/\b +
\d'\c'_{R}/\b'$ is the partition function
incorporating the three distinguishable pore
occupancy states  of the {\it entire} pore-$(R)$
reservoir ensemble.  Equation (\ref{SIGMAEFF}) is recovered
from (\ref{SIGMAAPPROX1}).  

While a single occupancy pore may not be molecularly
realistic, many of its qualitative physical characteristics
remain relevant for longer, multi-occupation pores. We
explicitly compute flows in two (Appendix B) and three
sectioned pores by calculating the coupled $9\times 9$ and
$27\times 27$ rate equations.  For these longer pores, the
reduction of $J_{V} < J_{max}$ due to decreased solvent
flux will also depend on where along the pore the membrane
first becomes impenetrable to solute.

%
% ******* Figure 3, Effective Reflection Coeffs **********
%
\begin{figure}
\begin{center}
\leavevmode
\epsfysize=4.0in
\epsfbox{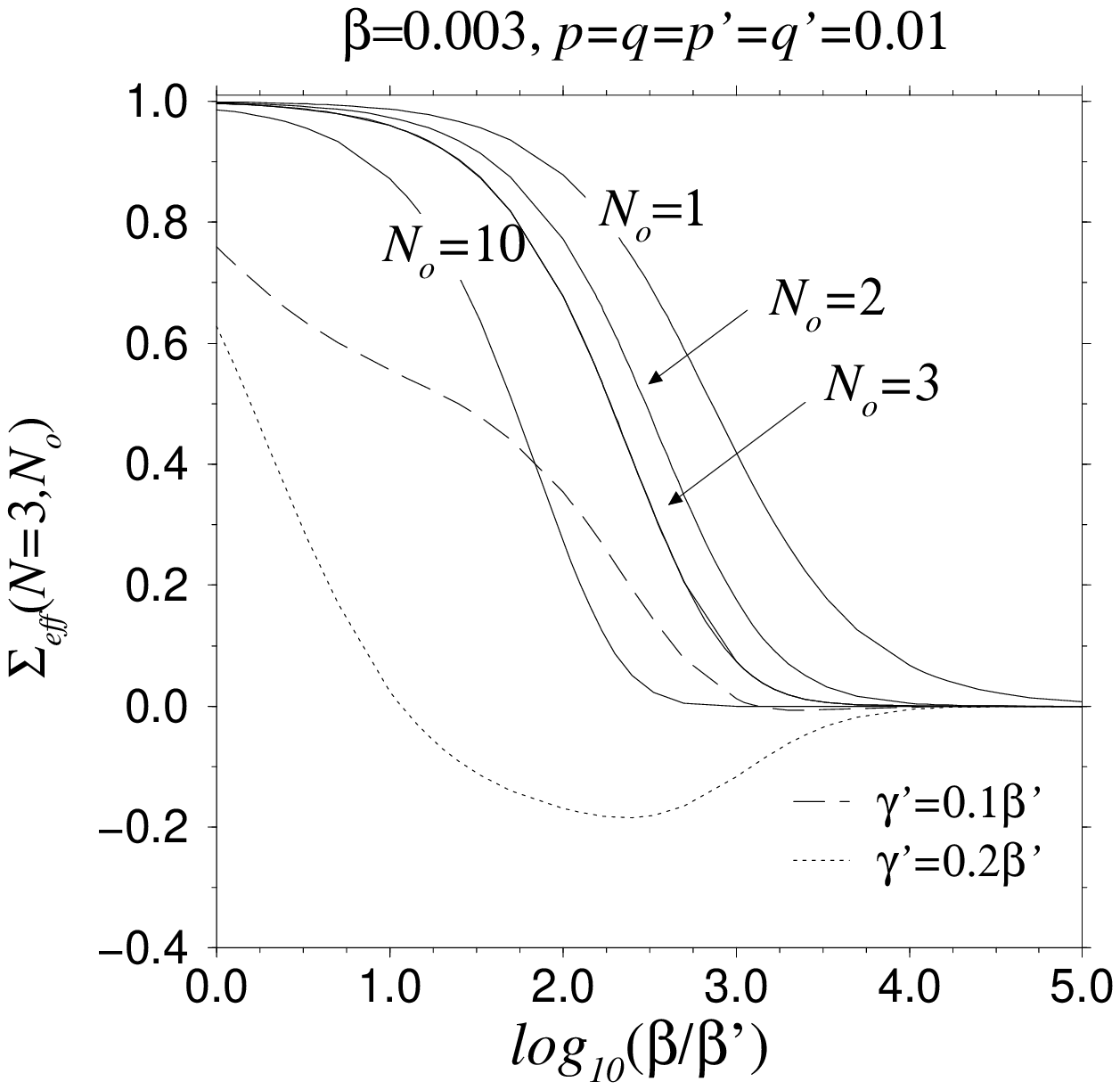}
\epsfxsize=2.7in
\epsfbox{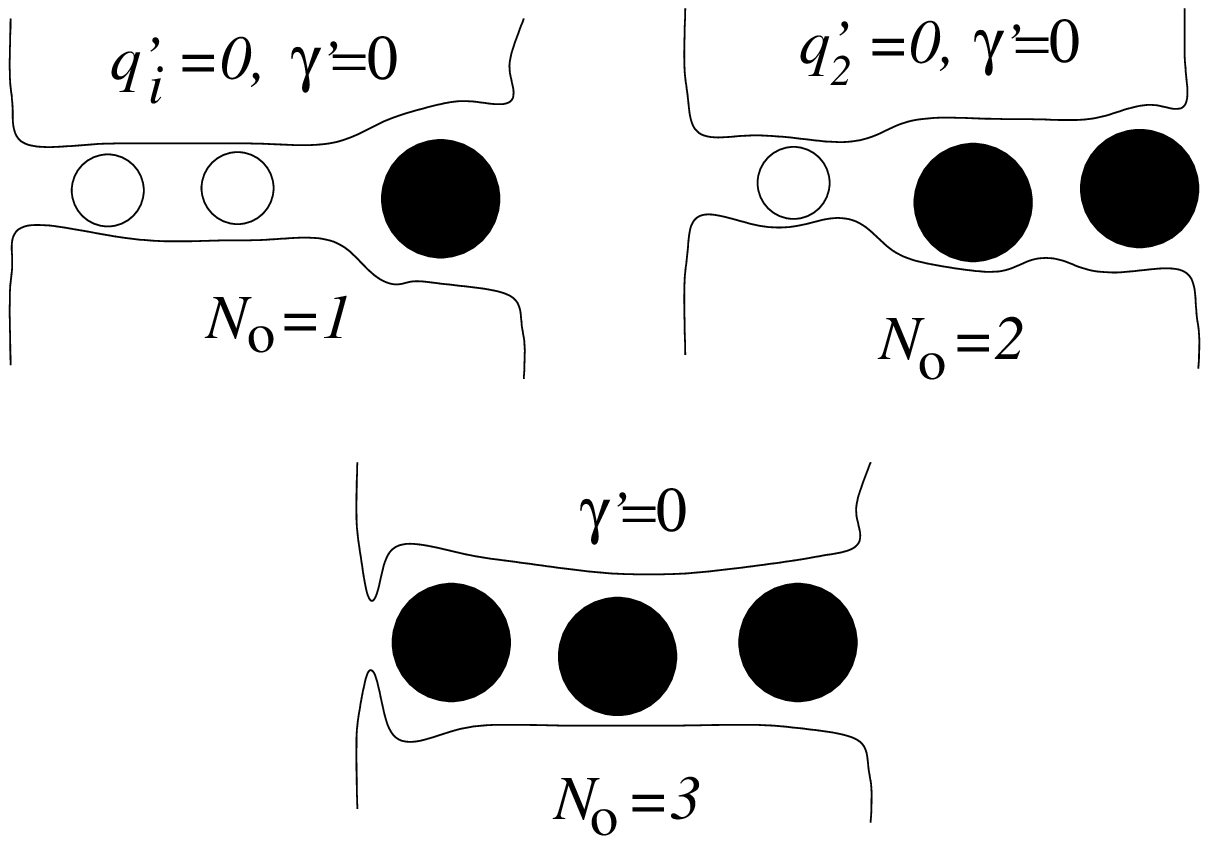} 
\end{center}
\caption{(a). Effective reflection coefficients
$\Sigma_{\em eff}\equiv J(N=3)/J_{max}(N=3)$, as a
function of solute-pore attraction $\b'$, for solute
impermeable pores ($\g'=0$, solid curves).  The
choice $\a=\d=\d'=0.01, \b=0.003$ corresponds to a
slightly solvent attracting pore. The three
sectioned pore allows solute entry into $N_{o} = 1,
2,\,\mbox{and}\, 3$ of its sections, schematically
shown in (b).  Also shown by the broken curves are
$\Sigma_{\em eff}$ (this does not include the volume
transfer by backflow of solute, treated in Section III)
when the solute is allowed to slowly permeate the
pore, $(\g'/\b' = 0.1, 0.2)$.  The curve labelled
$N_{o}=10$ is the equilibrium approximation
(\ref{EQUILN}) for a pore that allows solute entry
into $\sim 10$ sites.  Note the accuracy of
(\ref{EQUILN}) for $N_{o} = 3$, which is nearly
indistinguishable from the exact result. (b) 
Schematic of asymmetric pores allowing varying degrees of 
solute entrance.} 
\label{FIG3}
\end{figure}
%
% ********************************************************

Although solvent flux further decreases as solute is allowed to
permeate deeper into the pore (allowing more solvent blocking
configurations) solute-pore binding energies influence solute
occupation exponentially.  The sensitivity to solute-pore binding
energy for various $N_{o}$ is demonstrated by Fig.  \ref{FIG3}(a),
which shows the effective reflection coefficients for a three
sectioned pore, $\Sigma_{\em eff}(N=3,N_{o})=J(N,N_{o})/J_{max}(N)$,
as a function of the relative pore-solvent and pore-solute affinities.
For concreteness we take $dt=10$fs, such that $(v_{T}/\ell)\sim
1$ps$^{-1}$ sets $p_{i}=q_{i}\simeq 0.01 \ll 1$.  Across pore sections
that allow solute passage, we also assume $p'=q'=0.01$.  These are
reasonable estimates when considering molecular length, mass, and
energy scales under ambient conditions.  Sections that do not allow
solute passage are defined by $p'=0$ or $q'=0$. The parameters used
are $s=\a'\c'_{L} = \g'=0,\, \a=\d=\d'=0.01$, and $\b=0.003$.  These
values correspond to a slightly solvent-attracting pore
($\bar{\a}\simeq 3.33$), with identical intrinsic entrance rates for
solvent or solute particles.  The concentration $\c'_{R} = 0.1/55.556$
corresponds to a 100mM aqueous solution in $(R)$. The measure of
relative affinity $\log_{10}(\b/\b')$ is varied by tuning the
solute-pore attraction $\b'$ from $\a'_{0}=0.01\rightarrow 0$.
$N_{o}\leq N$ is the number of sections into the pore the solute can
enter from $(R)$, as shown in Fig. \ref{FIG3}(b).  For example, when
$N_{o}=2$, $p'_{1}=p'_{2}=q'_{1}=0.01$ and $q'_{2}=0$.  The upper
three solid curves in Fig. \ref{FIG3} represent the flux reduction
from $J_{max}(N=3)$ due to solute entrance into one, two, and all
three sites, corresponding to $q'_{i}=0, q_{2}=0$, and $\g'=0$
respectively.

The decrease of $\Sigma_{\em eff}$ for larger $N_{o}$ (the number of
sites open or accessible to solute) can be physically understood in
terms of a multisite Langmuir adsorption isotherm similar to that
applied to the one-site model. The probability that there are no
solute particles in any of the $N_{o}$ accessible sites is

\begin{equation}
\Sigma_{\em eff}(N, N_{o})  
\simeq  \left[
{1+\d\c_{R}/\b \over
1+\d\c_{R}/\b+\d'\c'_{R}/\b'}\right]^{N_{o}}. 
\label{EQUILN}
\end{equation}

\noindent This Increasing $N_{o}$, enhances
the probability of nonconducting pores and can also be thought of as
an additional entropic factor $k_{B}T\ln N_{o}$ in free energy
difference favoring nonconducting (solute adsorbed pores) over
conducting states.  Equation (\ref{EQUILN}) is valid provided $J$ is
too small to significantly affect {\it equilibrium} occupancies.  For
the parameters used, the estimate (\ref{EQUILN}) for $\Sigma(N_{o}=3)$
is nearly indistinguishable from the exact solution.  The effective
reflection coefficient (\ref{EQUILN}) for $N\geq N_{o}=10$ ($\g'=0,
s=0$) is also indicated in Fig. \ref{FIG3}.  Note that the estimated
relative flow reduction, (\ref{EQUILN}), is independent of the total
pore length $N$.

%
% ********** Figure 4, conc. dependence of \Sigma **********
%
\begin{figure}
\begin{center}
\leavevmode
\epsfysize=3.4in
\epsfbox{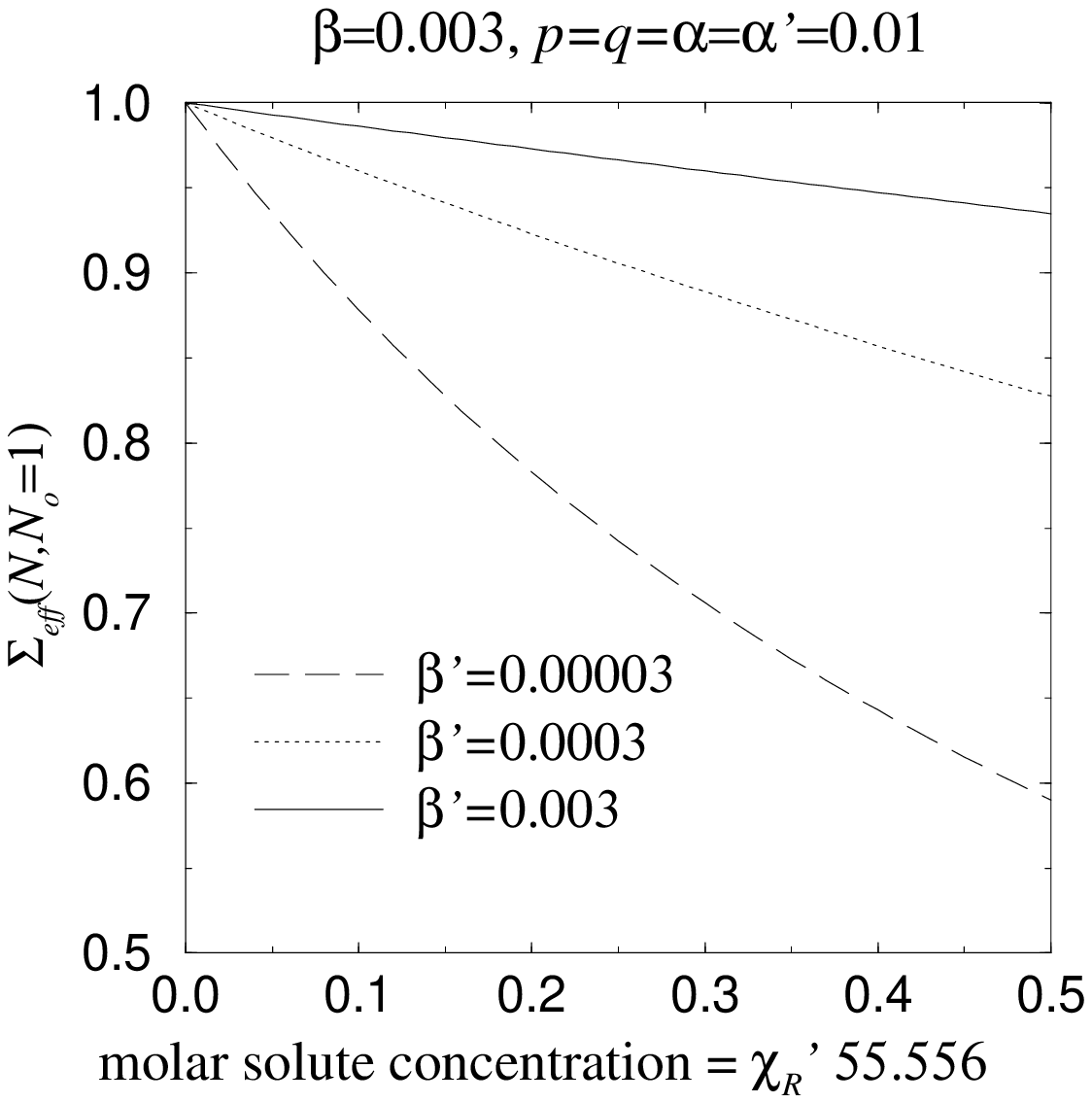}
\end{center}
\caption{Solute concentration dependences of 
$\Sigma_{\em eff}$ are strongest highly attracting 
solute-pore species (small $\b'$). The $N$
dependence is negligible with under these
parameters, consistent with the multisite 
Langmuir adsorption model (\ref{EQUILN}).}
\label{FIG4}
\end{figure}
%
% **********************************************************
%

Figure \ref{FIG4} shows the dependence of $\Sigma_{\em eff}(N,
N_{o}=1)$ on solute concentration for $N=1,2,3$ at various solute
bindings $\b'=0.00003, 0.0003, 0.003$.  All other parameters are
identical to those used in Fig.  \ref{FIG3}.  The reflection
coefficient decreases for larger solute concentration and smaller
$\b'$ as expected.  The independence of (\ref{EQUILN}) on $N$ is also
confirmed by the curves in Fig. \ref{FIG4}, which are each three
indistinguishable curves corresponding to $N=1, 2,$ and 3.

% **************************************************
% **************************************************

\subsection{Solute Permeable Pores}

When solute can steadily pass from $(R)$ to $(L)$ on
time scales of the measurement,\cite{PERMEABLE} the solute flow can
nonnegligibly contribute to total volume flow.  This
flow comprises the solvent flux tending in one direction
and solute flux tending in another.  
%In the single
%section model the sign of the total volume flow under
%isobaric conditions is determined by $J_{V}(N=1) \propto
%\a\g\g'\tv-\a'\g\g'\tv'$. 
Therefore, the measured
reflection coefficient $\Sigma$ (Eqn. (\ref{SIGMA}))
can even be negative. The solvent current 
$J$ can also be independently negative, {\it i.e.}  driven back by a
strong counterflowing solute.  Negative solvent flow is shown by the
broken curves in Fig.  \ref{FIG3} where the pore is permeable to
solute ($\g' >0$). The reflection coefficients measured and defined by
$\Sigma_{\em eff}$ or (\ref{SIGMA}) can thus be negative. However,
for the $N=1$ case, we see from (\ref{J1}) that increasing $\g'$
from zero can actually {\it increase solvent} flow $J$.  This occurs
particularly as $\c'_{L}\rightarrow 0$ and is the consequence of the
additional route for emptying the solute from the membrane site to
$(L)$, increasing the probability for a solute-free conducting state.
This ``backside'' exiting effect will only occur for very short pores
where a single kinetic step governed by $\g'$ (in addition to $\b'$)
renders the pore conducting and will disappear as $N$ increases.

The {\it total} volume flux, $J_{V}=\tv J +\tv' J'$ for
$N=1$ found from (\ref{J1}) and $J'$ can also {\it
increase} as $\g'$ increases provided $(\partial
J_{V}/\partial \g')>0$, which yields (for
$\c'_{L}=0$,$\a=\d$, and $\b=\g$)

\begin{equation}
{\b \over \b'} > {2(\a+\b)-\a\c'_{R}\over 
\a\c'_{R}(\tv/\tv')-2\d'\c'_{R}}
\end{equation}
 
\noindent Thus, a single site pore that is
impermeable ($\g' = 0$) but allows solute entry only
($\d'\neq 0$) can actually {\it increase} net volume
flow as it is made slightly permeable ($\g' > 0$).
This anomalous behavior is more prevalent when
$\tv'/\tv$ and $\c'_{L}$ are small, since the volume
backflow due to solute would be small and solute
re-entrance rates from $(L)$ vanish.

%\noindent The $N=1$ limit of a perfectly semipermeable (only allowing solvent
%into or through) is recovered as $\delta',\,\alpha'\rightarrow 0$.  

%Although the one-site model gives a simplified view of a complex biological
%pore, it contains all the ingredients necessary to qualitatively explain
%solute-dependent behavior and ultrafiltration of solute permeable membranes. In
%fact, the flux reduction effects of the one-site model is typically stronger
%than that of pores which allow multiple occupancy, as can be seen in Fig. 
%\ref{SIGMA2} for example. 

% *********** Fig. 5, various stopping pts. *************
%
\begin{figure}
\begin{center}
\leavevmode
\epsfysize=3.4in
\epsfbox{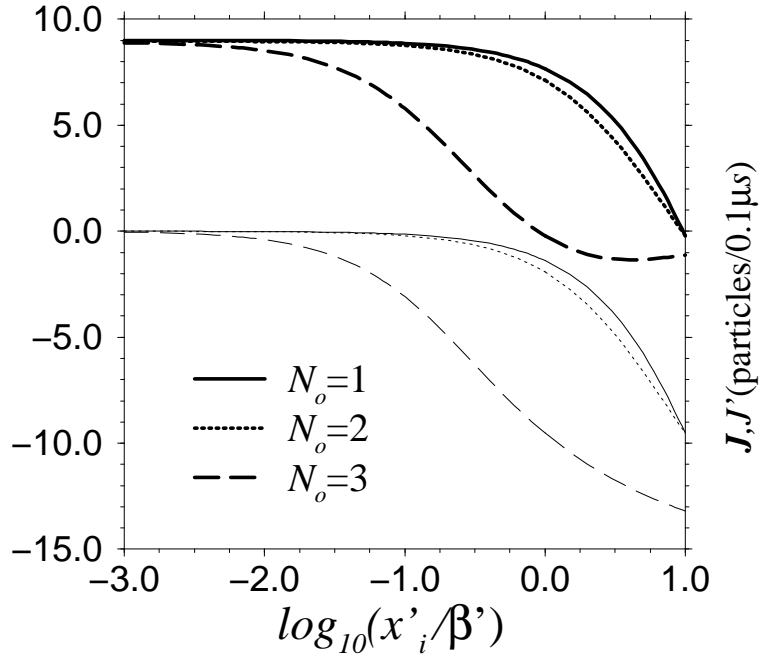}
\end{center}
\caption{Currents for various internal solute barrier
heights occuring at different positions. 
$N_{o}=1,2,3$ correspond to $1,
2, 3$ easily solute-accessible sites within an $N=3$ pore.  The
quantity $x'_{i}=q'_{1}, q'_{2}, \g'$ measures the
hopping rate to the left past this site.}
\label{FIG5}
\end{figure}
%
% *******************************************************

In Fig. \ref{FIG5}, we relax the impermeable solute constraint
and allow for passage of solute from $(R)\rightarrow (L)$.
Heights of the rate limiting internal barrier is varied
relative to $\b'$ at various internal positions $i$, where
$x'_{i} = q'_{1}, q'_{2}$, or $\g'$.  The
$x'_{i}/\b'\rightarrow 0$ limit of $J$ and $J'$ approach those
expected from simple osmosis resulting from an impermeable
(although pore-entering) solute, as studied in Fig.
\ref{FIG3}.  As impermeability is relaxed, $J$ decreases,
while $J'$ becomes negative.  The effects are most pronounced
for pores that have the largest number of easy entrance sites
({\it e.g.} $N_{o}=3$). In fact, for solute-binding pores where
$q'_{i}/\b' > 1$, and larger $N$, the negative solute flux can
actually drag, or pump solvent backwards ($N_{o}=3$ curves in
(b).  for $\g'/\b' \gtrsim 1$).

% ********* Fig. 6, Currents and occupations ***********
%
\begin{figure}
\begin{center}
\leavevmode
\epsfysize=3.8in
\epsfbox{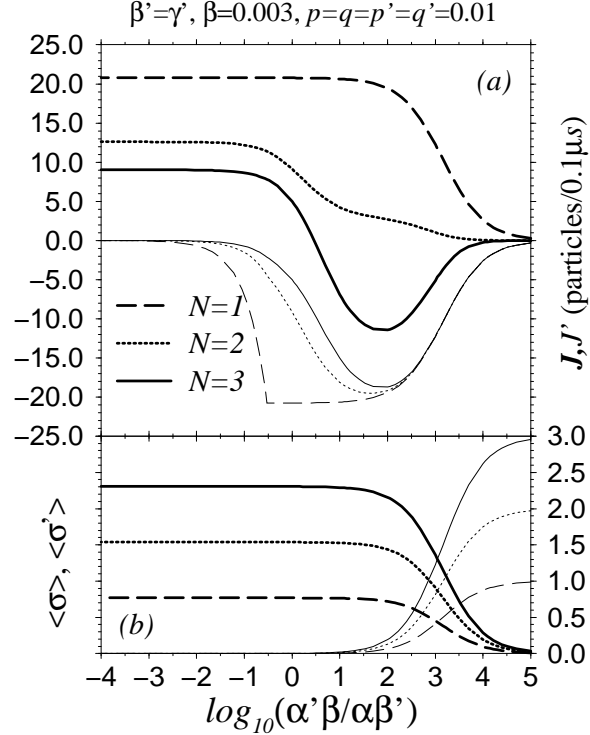}
\end{center}
\caption{(a). Solvent and solute fluxes as functions of
relative pore-solute, pore-solvent binding affinity
$\log (\a'\b/\a\b')$ (with $\a, \b$ fixed). To the
left of the discontinuity, (defined by
$\a'_{0}=0.003, \b'_{0}=0.01$ and most evident for
$N=1$) the pore is solute-repelling,
and $\a'$ is being varied, while to the
right, the pore is being made increasingly
solute-attractive by decreasing $\b'$.
(b). The associated total pore
occupancies.}
\label{FIG6}
\end{figure}
%
% *******************************************************

Negative solvent flux is shown in Fig. \ref{FIG6}.
Here, we assume a molecularly symmetric (with respect to both
solvent and solute) pore such that $\b'=\g', \b=\g$ and $q=p$
is constant along the pore, as assumed in the specific
dynamical model presented in Appendix A. Figure
\ref{FIG6}(a) shows the currents $J(N=3), J'(N=3)$ as
functions of the solute-pore affinity $\a'/\b'$ expressed in
terms of the logarithm of the relative (to solvent-pore binding
fixed at $\a=0.01, \b=0.003$) affinities. The discontinuity at
$\log(\a'_{0}/\b'_{0}) = \log(0.003/0.01)$ indicates the
crossover from solute-repelling to solute-attracting pores. 
Only at intermediate solute affinity, and longer pores ($N\geq
3$), does solute current $J'<0$ sweep solvent in a direction
opposite from that expected from simple osmosis. Pores
that repel solute rarely contain such particles that drive
solvent up its chemical potential gradient $\D \mu$. However,
pores that bind strongly to solute are choked off, and both $J$
and $J'$ diminish. Fig.  \ref{FIG6}(b) shows the associated total
pore occupancies $\sum_{i}^{N}\langle \s_{i}\rangle$  and
$\sum_{i}^{N}\langle \s'_{i}\rangle$.

% ********* Fig. 7, Concentration dependence ************
%
\begin{figure}
\begin{center}
\leavevmode
\epsfysize=5.4in
\epsfbox{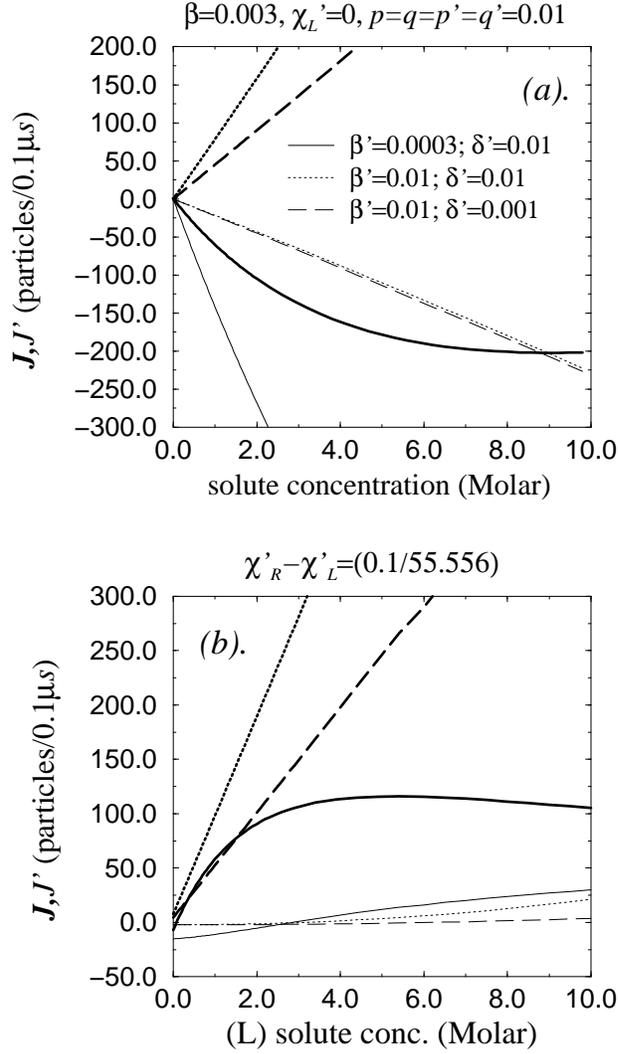}
\end{center}
\caption{Solvent and solute fluxes across symmetric
solute permeable pores as function of solute
concentrations. (a) $\c'_{L}=0$ as $\D \c' 
= \c'_{R}$ is varied. The lower three thin curves
and upper thick curves depict solute ($J' < 0$)
and solvent flow respectively. The parameters used
correspond to a solvent attracting pore with
strong solute binding (solid curves),
intermediate binding (dotted courves), and solute
repelling (dashed curves) pores. (b)
The currents as absolute concentration $\c'_{L}$ 
is varied, with $\D \c' = 0.1/55.556$ fixed.}
\label{FIG7}
\end{figure}
%
% *****************************************************

Figure \ref{FIG7}(a) shows the dependences of
$J(N=3)$ and $ J'(N=3)$ on solute concentration difference
$\D \c'$ with $\c'_{L}=0$. For 
strong solute-binding pores (low $\b'$, solid
curves), the solvent current $J(N=3)<0$ due to a
strong solute backflow $J'(N=3)<0$. In Fig
\ref{FIG7}(b), we vary $\c'_{L}$
while keeping $\D \c' = 0.1/55.556$ fixed. The
higher absolute solute concentrations in this case
decrease solvent occupation allowing a larger current
in the same direction as osmosis, and solute is 
swept in the same direction $J'>0$. However, for
too strong a solute binding, increasing solute
occupation eventually clogs the pore, precluding
efficient solvent flow (solid curves). Here, at low
absolute concentrations, $J, J'<0$ indicating that 
solute is pushing solvent back against the 
direction expected from simple osmosis.

% ************* Fig. 8,Slippage effects ****************
%
\begin{figure}
\begin{center}
\leavevmode
\epsfysize=3.0in
\epsfbox{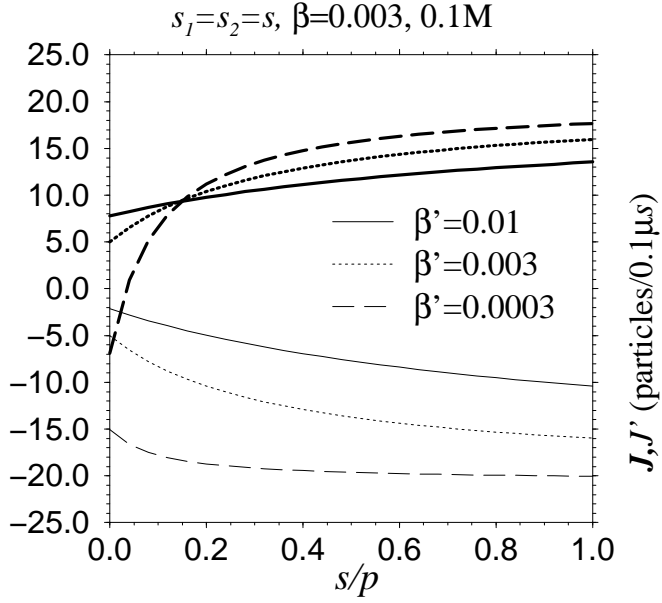}
\end{center}
\caption{Currents across a symmetric, $N=3$ pore
as a function of $s_{1}=s_{2}=s$. Concentrations
used are $\c'_{L}=0, \c'_{R}=0.1/55.556$.}
\label{FIG8}
\end{figure}
%
% ******************************************************

Effects of ``slip'' $s_{1}=s_{2}=s$ (see Fig.  \ref{FIG2}) are
explored in Fig. \ref{FIG8}.  The magnitudes of both $J(N=3)$ and
$J'(N=3)$ are increased as $s$ increases since the counterflowing
currents are occasionally allowed to more easily pass through each
other. This variation is strongest for smaller $\b'$ (strong
solute-binding pore) since large solute occupation configurations
would be most influenced by slip. Note the near symmetry between $J$
and $-J'$ when $\b=\b'=0.003$.

\section{Summary and Conclusions}

We have studied a commonly measured figure of merit, the reflection
coefficient $\Sigma$ (which describes the reduction in fluid volume
transport in the presence of permeable or membrane interacting
solutes). The single site model of Kalko {\it et
  al.}\cite{FISCHBARG95} is extended to include longer pores and the
effects of solute penetration and binding. The currents and reflection
coefficints (Figs. \ref{FIG3}-\ref{FIG8}) were computed in terms of
microscopic molecular parameters in some limiting, yet illustrative
cases.  Solutes that enter and bind to pore interiors block and reduce
solvent flux {\it without} being completely permeable, yielding an
effective reflection coefficient $\Sigma_{\em eff} < 1$.  Although
higher order corrections to van't Hoff's law have been studied
\cite{HILL,BPJ97} and is well understood in terms of virial expansions
of solute-solute interactions in the reservoirs, solute fouling arises
from higher order concentration terms resulting from solute-{\it
  membrane/pore} interactions. Although the phenomenological Onsager
coefficients $L_{P}, L_{PD}=L_{DP}, L_{D}$ can be qualitatively
approximated from our plots of $J$ and $J'$, care must be exercised
when interpreting measurements since solute interference effects can
be highly nonlinear, when Eqns. \ref{ONSAGER} do not apply.  The
magnitudes of the nonlinear effects depend on solute-pore interaction
energies as well as the number of configurations that are solvent
blocking, both of which can be inferred from microscopic models of the
particular pore-solute system being considered.  

For a one-site pore that allows solute entrance but remains
solute-impermeable ``solute interference'' nonetheless reduces the
solvent flux $J$ by a factor $\propto
\chi'_{R}\exp\left[(-E'_{\d'}+E'_{\b})/k_{B}T\right]$, where $\c'_{R}$
is the solute mole fraction, and $E'_{\d}$ and $E'_{\b}$ are the
solute-pore ``in'' and ``out'' energetic barriers.  Because the
nonlinearities are most pronounced for large $\d'/\b'$, {\it e.g.} a
very large $E'_{\b}$ or a highly solute-binding pore, strong solute
binding leads to the most effective solvent flow reduction.
Measurements of $J$ {\it vs.}  $\c'_{R}$ with different solutes should
have the same slope as $\c'_{R} \rightarrow 0$.  If different solutes
can indeed yield different slopes,\cite{HILLCOMM} the interactions are
too strong for the linear limit to have been reached. Extreme cases
are ``solutes'' such as mercurial compounds $p$-chloromecuribenzoate
(PCMB) that can reversibly destroy aquaporin pore function at very low
concentrations\cite{AQUAREV} due to strong binding; although here, the
binding is so strong that pore structure is presumably altered.
Experimentally, the solute concentration dependence on reflection
coefficients has been observed in various
systems.\cite{MCGANN,SOLOMON70,OPONG}

%Although typical membranes are
%much wider than one molecular diameter, biological pores likely only
%contain a few molecules at a time.\cite{FISCHBARG95} Therefore,
%$\ell$, and $q\simeq p$ can be rescaled such that longer pores can be
%qualitatively modeled by the few site ($N=2,3$) analyses presented
%here.

One possible experiment to probe solute blockage effects in asymmetric
(with respect to solutes) pores and flow rectification\cite{MCGANN} is
to compare the currents $J$ and $-J$ associated with solutes in the
$(L)$ and $(R)$ reservoirs.  For a symmetric pore, the magnitudes of
the currents will be equal.  However, for asymmetric pores, flow
induced by solutes in $(L)$ will be different from that generated by
solutes in $(R)$, the difference being manifested in higher order
terms in solute concentration.

If the pore is permeable to solute $(J' \neq 0)$, and $\Sigma$ is
defined by measuring the total volume flux $J_{V} = \tv J + \tv' J'$,
a negative reflection coefficient $\Sigma < 0$ can even occur when the
counterflowing solute adds to a possibly counterflowing solvent, or
overcompensates for the co-flowing solvent.  For solute-permeable
pores, correlations between volume flow and solute size typically used
in macroscopic descriptions, when extended to the microscopic regime
considered here, are not straightforward since $J_{V}$ depends
directly on $\tv'$ as well as on the rate parameters $\{\xi'\}$.  The
notion of molecular size alone, based on van der Waal's radii for
example, neglects pore-particle attractions and nonadditive
interactions among molecules such as hydrogen bonding.  For example,
urea may easily enter pores based on its small size (and thus have a
larger $N_{o}$), but participates in strong H-bonding in aqueous
solution, implying a very small $\a'/\b'$ (solute-repelling pore)
since H-bonds need to be broken prior to pore entry. Urea might then
be expected to yield a larger measured reflection coefficient than
that expected from its molecular size alone.\cite{TOON} H-bonding in
the bulk solution may also significantly change $D'$ and affect the
interpretation if unstirred layers are important.  Furthermore, the
additional solute interactions giving rise to $\Sigma_{\em eff},
\Sigma$ provide additional energy scales which can yield apparently
non-Arrhenius temperature dependences.\cite{BOEHLER}

The treatment given suggests that correlations between thermodynamic
solute-solvent parameters and pressure or osmosis driven flows can be
partly experimentally determined.  Specific volume changes of bulk
solute-solvent mixing, and enthalpies of mixing determine $\tv'/\tv$
and constrains $\{\xi,\xi'\}$.  For example, a solute with a negative
enthalpy of mixing (with solvent), $\Delta H_{m}$, would have a
smaller $\delta'/\beta'$ than a solute-solvent pair with larger or
positive $\Delta H_{m}$.  Equilibrium measurements of relative solvent
and solute absorption in pores also give relative measures of the
parameters $\d'/\b'$ and $\a/\b$. Further refinement of the parameters
$\{\xi\},\,\{\xi'\}$ may be obtained with more precise molecular
mechanics \cite{MILG2} or molecular dynamics dynamics (MD) using
appropriate molecular force fields.\cite{FISCHBARG95} Direct MD
simulation of osmosis across very thin membrane pores (single atomic
layers) have even been performed for high solute
concentrations.\cite{MURAD} Nonequilibrium MD simulations have also
been used to compute dynamic separation factors for two species
flows.\cite{TSOTSIS} Monte-Carlo simulations may also reveal
qualitatively interesting behavior in transport across longer
channels.

\acknowledgements

The author acknowledges D. Lohse and A. E. Hill for discussions and
comments, and T. J. Pedley and the reviewer for helpful suggestions.
This work was supported by The Wellcome Trust and a grant from The
National Science Foundation (DMS-9804870).

\begin{appendix}

\section{1D chain dynamics}

For arbitrarily long pores, the dynamical rules of particle transport
incorporating exclusion interactions can be formally implemented on a
discrete lattice.  If $\s_{i}(\s_{i}') \in \{0,1\}$ denote the
solvent(solute) particle occupation at site $i$, the instantaneous
rules for the microscopic transport of the two species in the pore
interior are mathematically defined by

\begin{equation}
\begin{array}{rl}
J_{i,i+1}  & =
\hat{p}\s_{i}(1-\s_{i+1})(1-\s'_{i+1})-\hat{q}
\s_{i+1}(1-\s_{i})(1-\s'_{i}) \\[13pt]
J'_{i,i+1} & =
\hat{p}'\s'_{i}(1-\s_{i+1})(1-\s'_{i+1})-
\hat{q}'\s'_{i+1}
(1-\s_{i})(1-\s'_{i})
\label{JINTERNAL}
\end{array}
\end{equation}

\noindent where $(\hat{p}, \hat{q}, \hat{p}', \hat{q}')
\in (0,1)$. The time variable has been suppressed for
notational simplicity.  These rules incorporate 
excluded volume into the dynamics of both solvent and
solute. Equations (\ref{JINTERNAL})
are supplemented with the boundary transition rates

\begin{equation}
\begin{array}{rl}
J_{(L),1} & =
\hat{\a}(1-\s_{1})(1-\s'_{1})-\hat{\g}\s_{1} \\[13pt]
J_{N, (R)} & =
\hat{\b}\s_{N}-\hat{\d}(1-\s_{N})(1-\s'_{N}) \\[13pt]
J'_{(L),1} & =
\hat{\a}'(1-\s_{1})(1-\s'_{1})-\hat{\g}'\s'_{1} \\[13pt]
J'_{N,(R)} & =
\hat{\b}'\s'_{N}-\hat{\d}'(1-\s_{N})(1-\s'_{N})
\label{JEXTERNAL} 
\end{array}
\end{equation}

Upon taking the time or ensemble average of the above equations, and
assuming the rate parameters are independent of the dynamical
variables $\s, \s'$, we formally obtain steady state flows that
depend on $\{\xi, \xi'\}$, the $dt\rightarrow 0$ limit of the averages
of $\{\hat{\xi}, \hat{\xi}'\}$.  However, the time averaged currents
are not easily solved due to higher order correlations among $\s, \s'$
such as $\langle \s_{i+1}\s'_{i} \rangle$, etc.\cite{ASEM,HAHN97}
Nonetheless, transport across very short (small $N$) pores can be
solved using rate equations defined by the rates $\{\xi\}, \{\xi'\}$
found from the mean of the distribution of $\{\xi\}, \{\xi'\} \in
\{0,1\}$.

\section{Two Species Two Site model}

Consider a membrane thick enough to accommodate a pore that can
simultaneously fit two particles, solvent or solute, along its axis. We
assume that the particle-particle interactions are dominated by
excluded volume effects such that longer ranged microscopic
interactions that cooperatively affect the particle hopping rates can
be neglected. At any instant, either of the two partitions in the pore
must be in one of three occupation states, empty, solvent(open circle),
or solute(filled circle).  Therefore, there are $3\times 3 = 9$
microscopic states denoted by

\begin{equation}
\begin{array}{llllll}
\: & \: & P_{1} \quad\quad\quad\quad &  \: 
& ^{\mbox{\circle*{13}}} & P_{5} \\[13pt]
\bigcirc & \: \quad & P_{2} \quad\quad\quad\quad &
^{\mbox{\circle*{13}}} & \: & P_{6} \\[13pt]
\: & \bigcirc & P_{3} \quad\quad\quad\quad  &
\!\!\!\bigcirc & ^{\mbox{\circle*{13}}} & P_{7}\\[13pt]
\bigcirc & \bigcirc & P_{4} \quad\quad\quad\quad &
^{\mbox{\circle*{13}}} & \!\!\! \bigcirc & P_{8}\\[13pt]
\: &\: & \: & ^{\mbox{\circle*{13}}} & ^{\mbox{\circle*{13}}} & P_{9}
\end{array}
\label{P9}
\end{equation}

\noindent where the transitions among the probabilities are
given by the coupled rate equations

\begin{equation}
\partial_{t}P_{i} + A_{ij}P_{j} = 0
\label{MASTER9}
\end{equation}

\noindent with the transition matrix defined by
${\bf A} = $

\begin{equation}
\!\!\!\!\!\!\!\!\!\!\!\!\!\!\left(
\begin{array}{ccccccccc}
\alpha_{T}+\delta_{T} & -\gamma & 
-\beta & 0 & -\gamma' &
-\beta' & 0 & 0 & 0 \\
-\alpha\c_L & \gamma+\delta_{T}+q & 
-p & -\beta & 0 & 0 & 0 & -\beta' & 0 \\
-\delta\c_R & -q & \alpha_{T}+\beta+p & 
-\gamma & 0 & 0 & -\gamma' & 0 & 0 \\
0 & -\delta\c_R & -\alpha\c_L & \gamma+\beta 
& 0 & 0 & 0 & 0 & 0 \\
-\alpha'\c'_L & 0 & 0 & 0 & 
\delta_{T}+\gamma'+q' & -p' 
& -\beta & 0 & -\beta' \\
-\delta'\c'_R & 0 & 0 & 0 & -q' &
\alpha_{T}+\beta'+p' & 0 & 
-\gamma & -\gamma' \\
0 & 0 & -\alpha'\c'_L & 0 & 
-\delta\c_R & 0 & \gamma'+\beta+s' & -s & 0 \\
0 & -\delta'\c'_R & 0 & 0 & 0 & 
-\alpha\c_L & -s' & +\gamma+\beta'+s & 0 \\
0 & 0 & 0 & 0 & -\delta'\c'_R & 
-\alpha'\c'_L & 0 & 0 & \beta'+\gamma' 
\end{array} \right)
\label{MATRIX9}
\end{equation}

\noindent where $\alpha_{T}(\delta_{T})\equiv
\alpha\c_L +\alpha'\c'_L(\delta\c_R +\delta'\c'_R)$.  The coefficients
are defined by the elementary steps shown in Figure \ref{FIG1};
$\alpha, \beta, \gamma, \delta$ and their primed analogues carry the
same physical meaning as in the one site model. The quantities $q(p),
q'(p')$, and $s(s')$ define the microscopic rates of transitions
$P_{2}\rightarrow P_{3}(P_{3}\rightarrow P_{2})$, $P_{5}\rightarrow
P_{6}(P_{6}\rightarrow P_{5})$, and $P_{8}\rightarrow
P_{7}(P_{7}\rightarrow P_{8})$, respectively. Probability
distributions $P_{i}$ are found by solving Eq. (\ref{MASTER9}) with
(\ref{MATRIX9}) and the constraint $\sum_{i=1}^{9}P_{i} = 1$. Form
these, the steady state solvent and solute currents are

\begin{equation}
\begin{array}{rl}
J & = \a\c_{L}(P_{1}+P_{3}+P_{6})-\g(P_{2}+P_{4}+P_{8}) \\[13pt]
J' & = \a'\c'_{L}(P_{2}+P_{3}+P_{6})-\g'(P_{5}+P_{7}+P_{9}).
\end{array}
\end{equation}

\noindent An analogous set of equations hold for the
kinetics of a three section, two-species pore.

\section{$N$ site, Two-Species Case}

We generalize the single section model to $N$ sections. Under
special cases, analytic expressions can be found for general
$N$.  Using notation describing the one section, the time
averaged fluxes of $\s$ and $\s'$ particles are

\begin{equation}
\begin{array}{l}
J = p\langle\s_{i}-\s_{i+1}\rangle+(p-s)
\langle\s_{i+1}\s'_{i}-\s_{i}\s'_{i+1}\rangle
\\[13pt]
J' =
p'\langle\s'_{i}-\s'_{i+1}\rangle+(p'-s)
\langle\s_{i}\s'_{i+1}-\s_{i+1}\s'_{i}\rangle,
\end{array}
\label{2FLUX}
\end{equation}

\noindent since $\langle\s_{i}\s_{i}' \ldots \rangle
= 0$. The steady state volume flux $J_{V} \equiv
\tilde{v}J+\tilde{v}'J'$ in the interior and 
at the boundaries are 

\begin{equation}
\begin{array}{l}
J_{V,i} = \tv p\langle\s_{i}-\s_{i+1}\rangle +
\tv' p'\langle \s_{i}'-\s_{i+1}'\rangle 
+\left[(p-s)\tv-(p'-s)\tv'\right]\langle
\s_{i+1}\s_{i}'-\s_{i}\s_{i+1}'\rangle \\[13pt]
J_{V,1} =
(\tv\a\c_{L}+\tv'\a'\c'_{L})
-Q_{L}\langle\s_{1}\rangle-Q'_{L}
\langle \s'_{1}\rangle \\[13pt]
J_{V,N} =
-(\tv\d\c_{R}+\tv'\d'\c'_{R})+
Q_{R}\langle\s_{N}\rangle+Q'_{R}
\langle\s'_{N}\rangle.
\end{array}
\label{2BOUNDARY}
\end{equation}

\noindent where 

\begin{equation}
\begin{array}{rl}
Q_{L} & = \tv\a\c_{L}+\tv'\a'\c'_{L}+\tv\g \\[13pt]
Q'_{L} & = \tv\a\c_{L}+\tv'\a'\c'_{L}+\tv'\g' \\[13pt]
Q_{R} & = \tv\d\c_{R}+\tv'\d'\c'_{R}+\tv\b \\[13pt]
Q'_{R} & = \tv\d\c_{R}+\tv'\d'\c'_{R}+\tv'\b'.
\end{array}
\end{equation}

\noindent The quadratic terms in $J_{V,i}$ cancel provided 

\begin{equation}
(p-s)\tv = (p'-s)\tv'.
\label{2CONDITION1}
\end{equation}

\noindent Under this special condition, it is useful
to define $\sigma^{+}_{i} \equiv p\tv
\s_{i}+p'\tv'\s'_{i}$ so that the current can be
summed as

\begin{equation}
J_{V} = {\langle \sigma^{+}_{1}-
\sigma^{+}_{N}\rangle \over N-1}.
\label{2JV}
\end{equation}

\noindent The volume flux $J_{V}$ is completely
determined when the boundary currents depend
only upon $\sigma^{+}$, which is the case 
provided

\begin{equation}
{Q_{L}\over Q'_{L}} = 
{Q_{R}\over Q'_{R}} 
= {\tv p\over \tv' p'}. 
\label{2CONDITION2}
\end{equation}

\noindent The above relationships, along with
(\ref{2CONDITION1}) provide constraints among $\a',\,
\b',\, \g',\,\d'$ given $\a,\,\b,\,\g,\,\d$.  In the
simple case of no-pass pores, $s=0$, a simple scaling
exists;

\begin{equation}
p' = p\phi, \quad \b' = \b\phi, \quad 
\g' = \g \phi
\end{equation}

\noindent where $\phi \equiv \tv/\tv'$. These special
conditions imply larger (in the aqueous bulk phase)
solute are more sluggish in hopping between sections
and into and out of the pore, and are not
qualitatively unreasonable. In this illustrative case,
the volume flux is 

\begin{equation}
J_{V}(N) =
{\g(\tv\d-\tv'\d')\D\c'+\tv(\a\b-\g\d)\c'_{L}+
\tv'(\a'\b-\g\d')\c'_{L}\over 
(N-1)Q_{L}Q_{R} +p\tv(Q_{L}+Q_{R})},
\label{JSPECIAL} 
\end{equation}

\noindent an implicit function of $\D P$.
In the $\c'_{L}=0$, isobaric ($\D P
\propto \a\b-\g\d=0$) limit the sign of 
$J'$ is determined by $\tv\d-\tv'\d'$.

%\begin{equation}
%J_{V}(N) = {\g(\tv\d-\tv'\d')\c'_{R} \over
%(N-1)\tv^{2}(\a+\g)(\b+\d)+p\tv^{2}(\a+\b+\g+\d)
%+\tv\left((N-1)(\a+\g)+p\right)
%(\tv'\d'-\tv\d)\c'_{R}}.
%\end{equation}

%\noindent In this special case,
%whether $J_{V}$ is $<$ or $> 0$ depends
%solely on $\tv\d-\tv'\d'$.

\section{Onsager relations for one site model}

We have only explicitly considered osmotic flow
between isobaric reservoirs $(L)$ and $(R)$. 
However, when a hydraulic pressure is applied, the
enthalpies between the two baths will generally be
unequal, and two independent thermodynamic driving
forces may exist. For microscopically symmetric
pores ($\a=\d, \a'=\d'$), the differences $\D E$ and
$\D E'$ can be readily related to hydrostatic
pressure differences.  Expanding the solutions
(\ref{J1}) and $J'(N=1)$ about $\D\chi'=0$ and
$P_{L}=P_{R}$, we find

\begin{equation}
\begin{array}{rl}
J & = P_{os}\D \Pi - L_{p}\D P  \\[13pt]
J' & =  -P'_{os}\D \Pi - L'_{p}\D P
\end{array}
\end{equation}

\noindent where the coefficients in terms of the
microscopic kinetic parameters are

\begin{equation}
\begin{array}{rl}
\displaystyle P_{os} & \displaystyle = 
{\a\g\g' \over 2n_{T}k_{B}T
(\a\g'\c_{L}+\a'\g\c'_{L}+\g\g')} \\[13pt]
\displaystyle P'_{os} & \displaystyle = 
{\a'\g\g' \over  2n_{T}k_{B}T(\a\g'\c_{L}+
\a'\g\c'_{L}+\g\g')} \\[13pt]
\displaystyle L_{p} & \displaystyle = {\a\g\g'\c_{L}
\tv \over 2k_{B}T(\a\g'\c_{L}+\a'\g\c'_{L}
+\g\g')} \\[13pt]
\displaystyle L'_{p} & \displaystyle = 
{\a'\g\g' \c'_{L}\tv' \over 
2k_{B}T(\a\g'\c_{L}+\a'\g\c'_{L}+\g\g')},
\end{array}
\end{equation}

\noindent where we have assumed an
attractive pore and linearized the exit rates
using their Arrhenius forms according to 

\begin{equation}
\b \simeq \g + \left({\partial \beta\over
\partial P_{R}}\right)_{P_{R}=P_{L}}\!\!\!\!\D P
= \g -{\g \over k_{B}T}\left({\partial E_{\b} \over
\partial P_{R}}\right)\D P,
\end{equation}

\noindent and the Maxwell relation $(\partial
E_{\b}/\partial P_{R})=\tv$.

Upon forming the conjugate flows and using
the definition (\ref{ONSAGER}), 

\begin{equation}
\begin{array}{rl}
\displaystyle L_{P} & \displaystyle =
-{\a\g\g'\c_{L}\tv^{2}+\a'\g\g'\c'_{L}\tv'^{2}
\over 2k_{B}T(\a\g'\c_{L}+\a'\g\c'_{L}+\g\g')}\\[13pt]
\displaystyle L_{PD}=L_{DP} & \displaystyle = 
{\a\g\g' \tv -\a'\g\g'\tv' \over 2n_{T}k_{B}T
(\a\g'\c_{L}+\a'\g\c'_{L}+\g\g')} \\[13pt]
\displaystyle L_{D} & \displaystyle  =
{\a\g\g'\c_{L}-\a'\g\g'\c'_{L} \over 
2n_{T}^{2}k_{B}T
(\a\g'\c_{L}+\a'\g\c'_{L}+\g\g')}
\end{array}
\end{equation}

\noindent which manifestly satisfies Onsager
reciprocity. Corresponding expressions for 
longer pores are very cumbersome but can be
evaluated with {\it e.g.} Mathematica.

\end{appendix}


\begin{references}

  
\bibitem{SCIENCE} D. A. Doyle, J. M. Cabral, R. A. Pfuetzner, A. L.
  Kuo, J. M. Gulbis, S. L. Cohen, B. T. Chait, and R. MacKinnon,
  Science, {\bf 280}, 69-77, (1998).
  
\bibitem{MITRA97} A. Cheng, A. N. van Hoek, M. Yeager, A. S. Verkmen,
  and A. K., and Mitra, Nature, {\bf 387}, 627-630, (1997).
  
\bibitem{JAP} B. K. Jap and H. Li, J. Mol. Biol., {\bf 251}, 413-420,
  (1995).

%\bibitem{ALBERTS} B. Alberts, D. Bray, J. Lewis  {\it
%Molecular Biology of the Cell,} (Garland, 1994) 

\bibitem{FINK} A. Finkelstein, {\it Water Movement Through Lipid
    Bilayers, Pore, and Plasma Membranes}, (Wiley-Interscience, New
  York, 1987).
  
\bibitem{ZEUTHEN} T. Zeuthen, Int. Rev. Cytology, {\bf 160}, 99,
  (1995).
  
\bibitem{KAT1} A. Katchalsky, P. F. and Curran, {\it Nonequilibrium
    Thermodynamics in Biophysics,} (Harvard University Press,
  Cambridge, 1965)

\bibitem{SU} A. Su, S. Mager, S. L. Mayo, and H. A.
Lester, Biophys. J., {\bf 70}, 762, (1996).
%A Multi-Substrate Single-File Model for Ion-Coupled
%Transporters.

\bibitem{2SITES}  K.-W. Wang, S. Tripathi, and S. B. Hladky,
%Ion Binding Constants for Gramicidin A
%Obtained from Water Permeability Measurements, 
J. Membr. Bio., {\bf 143}, 247, (1995).

\bibitem{MILG1} H. van Beijeren, K. W. Kehr, and R. Kutner, Phys. Rev.
  B, {\bf 28}, 5711, (1983).
  
\bibitem{ASEM} R. B. Stinchcombe and G. M. Sch\"{u}ltz, Phys.  Rev.
  Lett., {\bf 75}, 140, (1995); {\it Nonequilibrium Statistical
    Mechanics in One Dimension,} Ed. V. Privman, (Cambridge University
  Press, 1997).

\bibitem{EVANS} M. R. Evans, D. P. Foster, C. Godr\`{e}che, 
and D. Mukamel, 
%``Asymmetric Exclusion Model with Two Species: 
%Spontaneous Symmetry Breaking,''
J. Stat. Phys., {\bf 80}, 69-102, (1995).

\bibitem{PRL} T. Chou, 
%``How fast can fluids squeeze through micro-pores,''
Phys. Rev. Lett., {\bf 80}, 85-88, (1998).

\bibitem{TOBE} T. Chou, To be published.

\bibitem{HAHN97} C. R\"{o}denbeck, J. K\"{a}rger, and
K. Hahn, 
%``Exact analytical description of tracer exchange 
%and particle conversion in {\it single-file} systems,'' 
Phys. Rev. E, {\bf 55}(5), 5697-5712,
(1997).

\bibitem{HILLREV} A. E. Hill, Int. Rev. Cytology, {\bf 163}, 1-42,
  (1995).

\bibitem{HILL} T. L. Hill, {\it An
Introduction to Statistical
Thermodynamics}, (Dover, New York, 1986).

\bibitem{SOKOL92} S. Sokolowski, 
%``Molecular dynamics studies of adsorption and
%flow of fluids through geometrically non-uniform 
%cylindrical pores,''
Molecular Physics, {\bf 75}(6), 1301, (1992).

%\bibitem{CUSSLER} E. L. Cussler, 
%{\it Diffusion: Mass Transfer in Fluid Systems,}
%(Cambridge University Press, Cambridge, 1984).

%\bibitem{GUELL} D. C. Guell and H. Brenner, ``Physical Mechanism
%of Membrane Osmotic Phenomena,'' Ind. Eng. Chem. Res., {\bf 35}, 3004-3014,
%(1996).

%\bibitem{CRACKNELL} R. F. Cracknell, D. Nicholson, and
%N. Quirke, 
%``Direct Molecular Dynamics Simulation of
%Flow Down a Chemical Potential 
%Gradient in a Slit-Shaped Pore,'' 
%Phys. Rev. Lett., {\bf 74}(13), 2463-2466,
%(1995). 

\bibitem{FOOT1} Since $\a$ is an intrinsic entrance, or ``on'' rate,
  and $\b$ is an ``off'' rate, the ratio $\bar{\a}\equiv \a/\b$ gives
  an effective equilibrium solvent-pore binding constant averaged
  across the pore.
  
\bibitem{MCGANN} C. J. Toupin, M. Le Maguer, and L. E.
McGann, Cryobiology, {\bf 26}, 431-444, (1989).

\bibitem{SOLOMON70} R. I. Sha'afi, G. T. Rich, D. C. Mikulecky,
and A. K. Solomon, J. Gen. Physiol.,
{\bf 55}, 427-450, (1970).

%\bibitem{GALEY} W. R. Galey and J. Brahm, 
%``The failure of hydrodynamic analysis to 
%define pore size in cell membranes,'' 
%Biochim. et Biophysica Acta, {\bf 818},
%425-428, (1985).

\bibitem{PERMEABLE} Strictly speaking, all membranes are solute
permeable, given sufficient time for particle tunneling.

%\bibitem{KOHLER} H.-H. Kohler and K. Heckmann,
%``Unidirectional fluxes in staurated single-file pores
%of biological and artificial membranes.  I. Pores
%containing no more than one vacancy,'' 
%J. Theor. Biol., {\bf 79}, 381-401, (1979).

%\bibitem{FORD95} D. M. Ford, and E. D. Glandt, 
%``Steric hindrance at the entrances
%to small pores,'' {\it J. Membr. Sci.,} {\bf 107}(1-2), 47-57, (1995).

\bibitem{FISCHBARG95} S. G. Kalko, J. A. Hernandez, 
%J. R. Grigera and J. Fischbarg, 
%``Osmotic permeability in a molecular dynamics
%simulation of water transport through a single-occupancy pore,'' 
Biochim. et Biophys. Acta-Biomembranes, {\bf 1240}(2),
159, (1995). 

\bibitem{BPJ97} J. A. Cohen and S. Highsmith, 
%``An Improved Fit to Website Osmotic Pressure Data,'' 
Biophys. J., {\bf 73}, 1689, (1997); C. Reid and
R. P. Rand, 
%``Fits to Osmotic Pressure Data,''
Biophys. J., {\bf 73}, 1692, (1997).

\bibitem{HILLCOMM} A. E. Hill, Private Communication 

%\bibitem{LOO96} Loo, D. D. F., Zeuthen, T., Chandy, G., 
%and Wright, E. M. 1996 
%``Cotransport of water by the Na$^{+}$/glucose cotransporter,'' 
%Proc. Natl. Acad. Sci., {\bf 93}, 13367-13370

\bibitem{AQUAREV} C. H. van Os, P. Deen, J. A. Dempster,
Biochim. Biophys. Acta, {\bf 1197}, 291, (1994).

\bibitem{OPONG} W. S. Opong and A. L. Zydney, 
%``Effect of membrane structure and protein 
%concentration on the osmotic reflection coefficient,'' 
J. Membr. Sci., {\bf 72}, 277, (1992).

%\bibitem{GRAM} E. Bamberg and P. L\"{a}uger, ``Channel
%formation kinetics of gramicidin A in lipid bilayer
%membranes,'' J. Membr. Biol., {\bf 11}, 177-194, (1973).

\bibitem{TOON} M. R. Toon and A. K. Solomon,
%``Permeability and Reflection Coefficients of Urea and
%Small Amides in the Human Red Cell,'' 
J.  Membrane Biol., {\bf 153}, 137, (1996).

\bibitem{BOEHLER} B. A. Boehler, J. de Gier, L.  L. M.
van Deenen, 
%``The Effect of Gramicidin A on the
%Temperature Dependence of Water Permeation Through
%Liposomal Membranes Prepared from Phosphatidylcholines
%with Different Chain Lengths,'' 
Biochimica et Biophysica Acta, {\bf 512}, 480,
(1978).

\bibitem{MILG2} D. Sholl and K. A. Fichthorn, 
J. Chem. Phys., {\bf 107}, 4384, (1997).

\bibitem{MURAD} S. Murad, P. Ravi, and J. G. Powles,
J. Chem. Phys., {\bf 98}(12), 9771, (1993).

\bibitem{TSOTSIS} L. Xu, M. G. Sedigh, M. Sahimi, 
and T. T. Tsotsis, Phys. Rev. Lett., {\bf 80},
3511, (1998). 


\end{references}
\end{document}